\documentclass[aps,amsmath]{revtex4}
\usepackage{graphicx}  
\usepackage{dcolumn}   
\usepackage{amssymb}   
\usepackage{rotating}  
\def\MET{{\mbox{$E\kern-0.57em\raise0.19ex\hbox{/}_{T}$}}}
\def\met{{\mbox{$E\kern-0.57em\raise0.19ex\hbox{/}_{T}$}}}
\def\DZ{D0 }
\def\DZero{D0 }

\def\lmet{$WH\rightarrow \ell\kern-0.45em\raise0.19ex\hbox{/} \nu b\bar{b}$}

\begin{document}

\rightline{FERMILAB-CONF-11-372-E}
\rightline{CDF Note 10608}
\rightline{\DZ Note 6230}
\vskip0.5in

\title{Combined CDF and \DZ upper limits on {\boldmath $gg\rightarrow H\rightarrow W^+W^-$}
and constraints on the Higgs boson mass in fourth-generation fermion models with up to
8.2 fb$^{-1}$ of data\\[2.5cm]}

\author{
The TEVNPH Working Group\footnote{The Tevatron
New-Phenomena and Higgs Working Group can be contacted at
TEVNPHWG@fnal.gov. More information can be found at http://tevnphwg.fnal.gov/.}
 }
\affiliation{\vskip0.3cm for the CDF and \DZ Collaborations\\ \vskip0.2cm
\today}
\begin{abstract}
\vskip0.3in
We combine results from searches by the CDF and D0 Collaborations for a standard model Higgs
boson ($H$) in the processes $gg\rightarrow H\rightarrow W^+W^-$ and
$gg\rightarrow H\rightarrow ZZ$ in $p{\bar{p}}$
collisions at the Fermilab Tevatron Collider at
$\sqrt{s}=1.96$~TeV.  With 8.2~fb$^{-1}$ of integrated luminosity analyzed at CDF and
8.1~fb$^{-1}$ at D0, the 95\% C.L. upper limit on $\sigma(gg\rightarrow H)\times \mathcal{B}(H\rightarrow W^+W^-)$
is 1.01 pb at $m_H=120$~GeV, 0.40 pb at $m_H=165$~GeV, and 0.47 pb at $m_H=200$~GeV.  Assuming
the presence of a fourth sequential generation of fermions with large masses,
 we exclude at the 95\% Confidence Level a standard-model-like Higgs boson with a mass between 124 and 286~GeV.
\\[2cm]
{\hspace*{5.5cm}\em Preliminary Results}
\end{abstract}

\maketitle

\newpage
\section{Introduction} 

Exploring the mechanism for the breaking of the $SU(2)\times U(1)$
electroweak gauge symmetry is a priority in high energy physics.  Not
only are this symmetry and its breaking~\cite{higgs} necessary
components for the consistency of the successful standard model
(SM)~\cite{gws}, but measurable properties of the breaking mechanism
are also very sensitive to possible phenomena that have not yet been
observed at collider experiments.  Measuring these properties, or
setting limits on them, can constrain broad classes of extensions to
the SM.

A natural extension to the SM that can be tested with Higgs boson
search results at the Fermilab Tevatron Collider is the addition of a
fourth generation of fermions with masses much larger than those of
the three known generations~\cite{fourthgen}.  While fits to precision
electroweak data favor a low-mass Higgs boson in the SM, the addition
of a fourth generation of fermions to the SM modifies the fit
parameters such that a much heavier Higgs boson is 
allowed~\cite{He:2001tp,g4_hdecay}.  Currently available data
are consistent with a Higgs boson of mass 900~GeV with suitably chosen
fourth generation fermion masses~\cite{gfitter2011}.  Precision
measurements of the $Z$ boson decay width~\cite{lepzpole} exclude
models in which the fourth neutrino mass eigenstate has a mass less
than 45~GeV.  If the neutrino masses are very large, however, a fourth
generation of fermions is not excluded.

One consequence of the extra fermions is that the $ggH$ coupling is
enhanced by a factor of roughly three relative to the SM
coupling~\cite{arik,g4_hdecay,abf}.  The reason for this is that the
lowest-order $ggH$ coupling arises from a quark loop.  The top quark
contribution is the largest due to its large coupling with the Higgs
boson.  In the limit $m_{q4}\gg m_H$ where $m_{q4}$ is the
fourth-generation quark mass, the Higgs boson coupling cancels the
mass dependence for each of the three propagators in the loop, and the
contribution to the $ggH$ coupling becomes asymptotically independent
of the masses of the two fourth-generation quarks.  Each additional
fourth-generation quark then contributes as much as the top quark, and
the $ggH$ coupling is thus enhanced by a factor $K_e$ of approximately
three.

The production cross section will be enhanced by a factor of $K_e^2$.
For $m_H$ near the low end of our search range, $m_H \approx 110$~GeV,
the $gg\rightarrow H$ production cross section is enhanced by roughly
a factor of nine relative to the SM prediction.  This factor drops to
approximately 7.5 near the upper end of the search range, $m_H \approx
300$~GeV, assuming asymptotically large masses for the
fourth-generation quarks.  The reason for this drop is that the
denominator of the enhancement factor, the SM cross section, has a
larger contribution from the SM top quark as $m_H$ nears $2m_t$.
The partial decay width for $H\rightarrow gg$ is enhanced by the same
factor as the production cross section.  However, because the decay
$H\rightarrow gg$ is loop-mediated, the $H\rightarrow W^+W^-$ decay
continues to dominate for Higgs boson masses $m_H>135$~GeV.

We consider two scenarios for the masses of the fourth-generation
fermions.  In the first scenario, the ``low-mass'' scenario, we set
the mass of the fourth-generation neutrino $m_{\nu 4}=80$~GeV, and the
mass of the fourth-generation charged lepton $m_{\ell 4}=100$~GeV, in
order to evade experimental constraints~\cite{L3_lepton} and to have
the maximum impact on the Higgs boson decay branching ratios.  In the
second scenario, the ``high-mass'' scenario, we set $m_{\nu 4}=m_{\ell
4}=1$~TeV, so that the fourth-generation leptons do not affect the
decay branching ratios of the Higgs boson.  In both scenarios, we
choose the masses of the quarks to be those of the second scenario in
Ref.~\cite{abf}, that is, we set the mass of the fourth-generation
down-type quark to be $m_{d4}=400$~GeV and the mass of the
fourth-generation up-type quark to be $m_{u4}=m_{d4}+50~{\rm
GeV}+10\log\left(m_H/115~{\rm GeV}\right){\rm GeV}$.  The other mass
spectrum of Ref.~\cite{abf} chooses $m_{d4}=300$~GeV, resulting in
slightly larger predictions for $\sigma(gg\rightarrow H)$ and nearly
identical decay branching ratios.  We use the
next-to-next-to-leading order (NNLO) production cross section
calculation of Ref.~\cite{abf}, which builds on the NNLO SM
calculations of
Refs.~\cite{harlander1,melnikov1,ravindran1,anastasiou,bucherer1,spira1,nnloggh,aglietti}.

The CDF and D0 Collaborations have searched for the SM Higgs boson in
the decays $H\rightarrow W^+W^-$ and $H\rightarrow ZZ$ using all SM
production processes: $gg\rightarrow H$, $qq\rightarrow WH$,
$qq\rightarrow ZH$, and vector-boson fusion
(VBF)~\cite{cdfwwsum11,cdfzzsum11,dzHWW,dzHWWtau,dzHWWjj}.  The results of these
searches for the SM Higgs boson cannot be used directly to constrain
fourth-generation models, as the $ggH$ coupling is enhanced but the
$WWH$ and $ZZH$ couplings are not, and the signal acceptances and the
backgrounds in the multiple analysis channels differ for the various
production modes.

Previously, the CDF and D0 Collaborations have published searches for
the process $gg\rightarrow H\rightarrow W^+W^-$, also neglecting the
$WH$, $ZH$, and VBF signal contributions~\cite{cdfwwprl,d0wwprl,PRDRC}.
This paper is an update of Ref.~\cite{PRDRC}, which provides a fourth-generation
interpretation of the $gg\rightarrow H\rightarrow W^+W^-$ searches, and a model-independent
limit on $\sigma(gg\rightarrow H\rightarrow W^+W^-)$.
Here we update the searches detailed in Ref.~\cite{PRDRC} with those using 8.2~fb$^{-1}$ from CDF~\cite{cdfwwsum11}
and 8.1~fb$^{-1}$ from D0~\cite{dzHWW} and include additional channels~\cite{cdfzzsum11,dzHWWtau,dzHWWjj}.
Interpretations of SM Higgs boson searches within the context of a fourth generation of fermions have been
performed by ATLAS~\cite{atlas4g} and CMS~\cite{cms4g}, which exclude $140<m_H<185$~GeV and 
$144<m_H<207$~GeV, respectively.

  We present new limits on $\sigma(gg\rightarrow H)\times
\mathcal{B}(H\rightarrow W^+W^-)$ in which the $gg\rightarrow H$
production mechanism is considered as the unique signal source.  These
limits are compared to models for Higgs boson production in which the
$ggH$ coupling is enhanced by the presence of a single additional
generation of fermions.  In this comparison, the decay branching
ratios of the Higgs boson are also modified to reflect changes due to
the fourth generation relative to the SM prediction.  While the decays
of the heavy quarks and leptons may include $W$ bosons in the final
state, we do not include these as additional sources of signal.  The
branching ratios for $H\rightarrow W^+W^-$ are calculated using {\sc hdecay}~\cite{hdecay} 
modified to include fourth-generation
fermions~\cite{g4_hdecay}.  The modified Higgs branching ratio to
$W^+W^-$ is multiplied by the cross section~\cite{abf} to predict the
fourth-generation enhanced $gg\rightarrow H \rightarrow W^+W^-$
production rate.  We include as an approximation to the uncertainties
on the decay branching ratios the relative uncertainties predicted for the
SM branching ratios due to uncertainties in $m_b$, $m_c$, and $\alpha_s$
presented in Ref.~\cite{dblittlelhc}.  

The branching ratio for the decay $H\rightarrow ZZ$ is also considerable
at large masses, having a peak at around $m_H=135$~GeV and rising above
$m_H=180$~GeV.  In order to include a search for $gg\rightarrow H\rightarrow ZZ$,
we introduce the assumption that
$\mathcal{B}(H\rightarrow W^+W^-)/\mathcal{B}(H\rightarrow ZZ)$ remains as predicted
by the SM, even though the production cross section is enhanced.  This assumption
holds true in fourth-generation models as predicted by the fourth-generation
version of {\sc hdecay} mentioned above.  In setting limits on $\sigma(gg\rightarrow H)\times
\mathcal{B}(H\rightarrow W^+W^-)$, we include the process $gg\rightarrow H\rightarrow ZZ$
assuming that its signal yield scales in the same way as the $W^+W^-$ channel.

The event selections are similar for the corresponding CDF and D0
analyses.  Both collaborations select events with large \met~and two
oppositely charged, isolated leptons, targeting the $H\rightarrow
W^+W^-$ signal in which both $W$ bosons decay leptonically.
D0 selects events containing electrons and/or muons, dividing the data sample
into three final states: $e^+e^-$, $e^\pm \mu^\mp$, and $\mu^+\mu^-$. Each final state is further
subdivided according to the number of jets in the event: 0, 1, or 2 or more (``2+'') jets.
However, in the fourth-generation interpretation and the model-independent
$gg\rightarrow H$ limits, due to the low S/B in the 2-jet multiplicity bin,
only the 0 and 1-jet multiplicity bins were taken into account.
Decays involving tau leptons are included in two orthogonal ways. A dedicated
analysis ($\mu\tau_{had}$) using 7.3 fb$^{-1}$
of data studying the final state involving a muon and a hadronic tau decay, 
is included~\cite{dzHWWtau}. Final states involving other tau decays 
and mis-identified hadronic tau decays are included in the $e^+e^-$, $e^\pm \mu^\mp$, 
and $\mu^+\mu^-$ final state analyses. The
CDF analysis separates opposite-sign candidate events into five
non-overlapping channels.  Events are classified by their jet
multiplicity (0, 1, or $\ge$ 2), and the 0 and 1 jet channels are
further divided according to whether both leptons are in the central
part of the detector or if either lepton is in the forward part of the
detector. The presence of neutrinos in the final state prevents event-by-event
reconstruction of the Higgs boson mass and thus other variables are
used for separating the signal from the background.  For example, the
angle $\Delta\phi(\ell,\ell)$ in signal events is smaller on average
than that in background events, the missing transverse momentum is
larger, and the total transverse energy of the jets is lower. The CDF
analyses use neural-network (NN) outputs as the final discriminants; the
list of network inputs includes likelihood ratio
discriminant variables constructed from matrix-element
probabilities~\cite{cdfwwprl}. The D0 $e^+e^-$, $e^\pm \mu^\mp$, and $\mu^+\mu^-$ final state channels use
random forest boosted decision tree (BDT) outputs as the final discriminants while
the $\mu\tau_{had}$ channel uses neural networks. Both experiments use separate NNs and BDTs for the
different jet multiplicities to distinguish the $gg\rightarrow H$ signal from the backgrounds for each
of the test masses, which are separated by increments of 5 or 10 GeV.

CDF introduces a search for
$gg\rightarrow H\rightarrow ZZ\rightarrow \ell^+\ell^-\ell^{\prime+}\ell^{\prime-}$ using
8.2~fb$^{-1}$ of data~\cite{cdfzzsum11}.  The discriminant variable is the reconstructed mass of the four
leptons $m_{\ell\ell\ell\ell}$.  The version of this search used for the Standard Model interpretation
also includes the expected yield for $WH$, $ZH$, and $VBF$ production, but these production modes
are assumed not to contribute when interpreting the results in this paper.

D0 also includes  for the first time channels in which one of the $W$
bosons in the $H \rightarrow W^+W^-$ process decays leptonically and the other
decays hadronically~\cite{dzHWWjj}.  Electron and muon final states are studied separately,
each with 5.3 fb$^{-1}$ of data.  Random forests are used for the final
discriminants.

Both CDF and D0 conduct all searches included in this paper in the range
$110<m_H<300$~GeV. The details of the signal and background estimations and the
systematic uncertainties are provided in
Refs.~\cite{cdfwwsum11,cdfzzsum11,dzHWW,dzHWWtau,dzHWWjj}.  We set limits on
$\sigma(gg\rightarrow H)\times \mathcal{B}(H\rightarrow W^+W^-)$ as a
function of $m_H$.
 We use the same two statistical methods employed in
Ref.~\cite{tevwwprl}, namely the modified frequentist (CL$_{\rm s}$)
and Bayesian techniques, in order to study the consistency of the
results.  Each method is applied at each test mass to calculate an
observed upper limit on $\sigma(gg\rightarrow H)\times
\mathcal{B}(H\rightarrow W^+W^-)$.  Pseudo-experiments drawn from
systematically varied background-only predictions are used to compute
the limits we expect to obtain in the absence of a signal. 

Correlated systematic
uncertainties are treated in the same way as they are in
Ref.~\cite{tevwwprl}.  The sources of correlated uncertainty between
CDF and D0 are the total inelastic $p\bar{p}$ cross section used in
the luminosity measurement, the SM diboson background production cross
sections ($WW$, $WZ$, and $ZZ$), and the $t{\bar{t}}$ and single top
quark production cross sections.
Instrumental effects such as trigger efficiencies, lepton
identification efficiencies and misidentification rates, and the jet
energy scales used by CDF and D0 remain uncorrelated.  To minimize the
degrading effects of systematics on the search sensitivity, the signal
and background contributions are fit to the data observations by
maximizing a likelihood function over the systematic uncertainties for
both the background-only and signal+background
hypotheses~\cite{fitting}.  When setting limits on
$\sigma(gg\rightarrow H)\times \mathcal{B}(H\rightarrow W^+W^-)$, we
do not include the theoretical uncertainty on the prediction of
$\sigma(gg\rightarrow H)\times \mathcal{B}(H\rightarrow W^+W^-)$ in
the fourth-generation models since these limits are independent of the
predictions.  When setting limits on $m_H$ in the context of
fourth-generation models, however, we include the uncertainties on the
theoretical predictions as described below.

Since both the CDF and the D0 analyses categorize events into separate
channels based on the number of reconstructed jets, we include the systematic
uncertainties on each jet category according to the procedure recommended
by Ref.~\cite{bnlaccord}.  This procedure treats the factorization and renormalization
scale uncertainties as independent in the inclusive $gg\rightarrow H$,
$gg\rightarrow H+1$ or more jets, and $gg\rightarrow H+2$ or more jets calculations.
We propagate these uncertainties through to the calculations of the exclusive
$H+0$~jets, $H+1$~jet, and $H+2$ or more jets calculations required to predict the
yields in the fourth-generation models.  When setting a limit on
$\sigma(gg\rightarrow H\rightarrow W^+W^-)$, we do not include a theoretical uncertainty
on the prediction of the quantity that we are constraining experimentally.  Nonetheless,
we do require theoretical input for the differential distributions of the number of jets
in $gg\rightarrow H$ production.  We therefore use only the factorization and renormalization
scale uncertainties on the $H+1$ or more jets and $H+2$ or more jets inclusive calculations
when setting a limit on the inclusive process.  We follow the PDF4LHC prescription~\cite{pdf4lhc} for
evaluating the parton distribution
function (PDF) uncertainties on the signal predictions.  QCD scale uncertainties that
affect the PDF predictions are included as part of the total scale uncertainty, added linearly
to the other components of the scale uncertainty.  The remaining component of the PDF uncertainty
is treated as uncorrelated with the QCD scale uncertainty.

The scale choice affects the $p_T$ spectrum of the Higgs boson when
produced in gluon-gluon fusion, and this effect changes the acceptance
of the selection requirements and also the shapes of the distributions
of the final discriminants.  The effect of the acceptance change is
included in the calculations of Ref.~\cite{anastasiouwebber} and
Ref.~\cite{campbell2j}, as the experimental requirements are simulated
in these calculations. The effects on the final discriminant shapes
are obtained by reweighting the $p_T$ spectrum of the Higgs boson
production in our Monte Carlo simulation to higher-order calculations.
The Monte Carlo signal simulation used by CDF and D0 is provided by
the LO generator {\sc pythia}~\cite{pythia} which includes a parton
shower and fragmentation and hadronization models.  We reweight the
Higgs boson $p_T$ spectra in our {\sc pythia} Monte Carlo samples to
that predicted by {\sc hqt}~\cite{hqt} when making predictions of
differential distributions of $gg\rightarrow H$ signal events. To
evaluate the impact of the scale uncertainty on our differential
spectra, we use the {\sc resbos}~\cite{resbos} generator, and apply
the scale-dependent differences in the Higgs boson $p_T$ spectrum to
the {\sc hqt} prediction, and propagate these to our final
discriminants as a systematic uncertainty on the shape, which is
included in the calculation of the limits.

\begin{figure}
 \begin{center}
\includegraphics[width=0.8\textwidth]{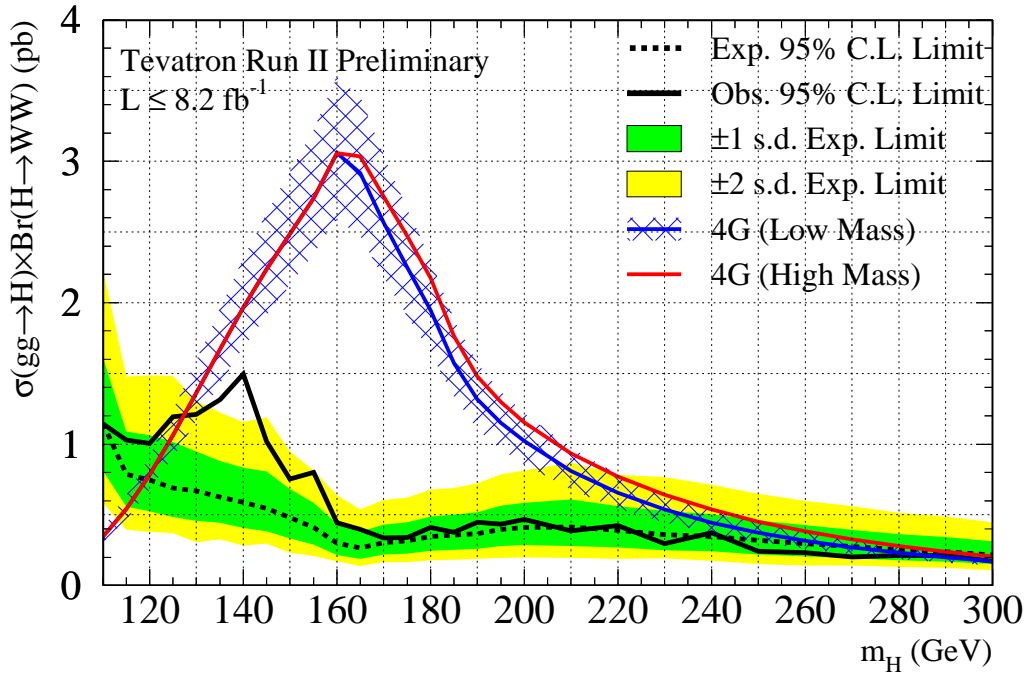}
 \end{center}
 \caption{
 \label{fig:xslimits}  The Tevatron combined  observed (solid black lines) and median expected (dashed
black lines) 95\% C.L. upper limits on $\sigma(gg\rightarrow H)\times \mathcal{B}(H\rightarrow W^+W^-)$.
 The shaded bands indicate the $\pm 1$~standard deviation (s.d.) and $\pm 2$~s.d. intervals
on the distribution of the
limits that are expected if a Higgs boson signal is not present.  Also shown the prediction
for a fourth-generation model in the low-mass and high-mass scenarios, 4G (Low mass) and 4G (High mass) respectively.
The hatched areas indicate the theoretical uncertainty from PDF and scale uncertainties.
The lighter curves show the high-mass theoretical prediction.
}
 \end{figure}

 The combined limits on $\sigma(gg\rightarrow H)\times
\mathcal{B}(H\rightarrow W^+W^-)$ are listed in Table~\ref{tab:limits}
for both the CL$_{\rm s}$ and the Bayesian methods, and are shown in
Fig.~\ref{fig:xslimits} along with the fourth-generation theory
predictions for the high-mass and low-mass scenarios.  The uncertainty
bands shown on the low-mass theoretical prediction are the quadrature
sum of the MSTW 2008 \cite{mstw1} 90\% C.L. PDF uncertainties and the factorization and renormalization
scale uncertainties from Table~1 of Ref.~\cite{abf}.  
The limits calculated with the two statistical methods agree
within 10\% for all Higgs boson mass hypotheses.  
An excess is seen in the neighborhood of
$m_H=140$~GeV.  This excess is localized to the D0 $H\rightarrow W^+W^-\rightarrow e^+\nu_e\mu^-{\bar{\nu}}_\mu$
search (including charge conjugate final states)~\cite{dzHWW}.  The excess is not as pronounced
in the SM-optimized search from D0~\cite{dzHWW}.  Combination with CDF, which sees no significant excess
at $m_H=140$~GeV~\cite{cdfwwsum11} results in a total local significance somewhat in excess of two standard deviations.
The trials factor, or look-elsewhere effect, further dilutes the significance, in this case by a large
factor due to the large range of $m_H$ explored.  We do not claim that this excess is statistically significant.

\begin{figure}
 \begin{center}
 \includegraphics[width=0.8\textwidth]{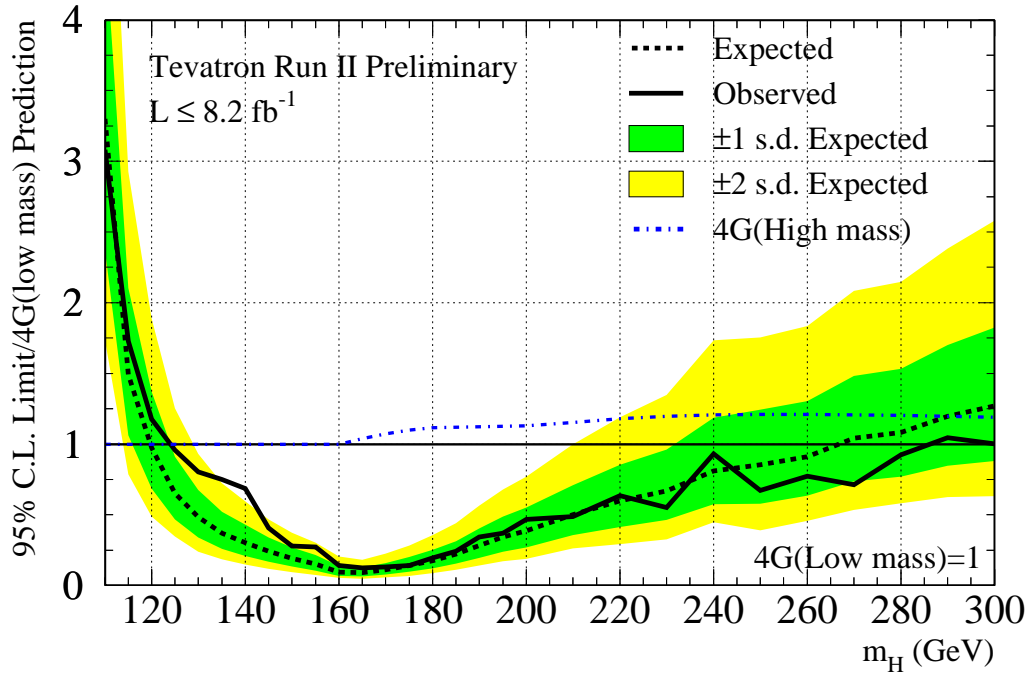}
 \end{center}
 \caption{
 \label{fig:4glimits}  The Tevatron combined  observed (solid black lines) and median expected (dashed
black lines) 95\% C.L. upper limits on the ratio of the Higgs boson production cross section to that
predicted by the 4G (Low mass) model; the uncertainties in the signal prediction
are included in the limit.  Also shown is the prediction of the signal rate
in the high-mass scenario, divided by that of the low-mass scenario.
}
 \end{figure}

In order to set limits on $m_H$ in these two scenarios, we perform a
second combination, including the uncertainties on the theoretical
predictions of $\sigma(gg\rightarrow H)\times \mathcal{B}(H\rightarrow
W^+W^-)$ due to scale and PDF uncertainties at each value of $m_H$
tested.  The resulting limits are computed relative to the model
prediction, and are shown in Fig.~\ref{fig:4glimits} for the low-mass
scenario, which gives the smaller excluded range of $m_H$.  In this
scenario, we exclude at the 95\% C.L. a SM-like Higgs boson with a
mass in the range 124 -- 286~GeV.  Using the median limits on
$\sigma(gg\rightarrow H)\times \mathcal{B}(H\rightarrow W^+W^-)$,
expected in the absence of a signal to quantify the sensitivity, we
expect to exclude the mass range 120 -- 267~GeV.  In the high-mass
scenario, which predicts a larger $\mathcal{B}(H\rightarrow W^+W^-)$
at high $m_H$ than that predicted in the low-mass scenario, we exclude
at the 95\% C.L. the mass range 124 -- 300~GeV and expect to exclude
the mass range 120 -- 290~GeV.  The upper edge of the observed
exclusion range in the high-mass scenario is determined by the fact
that we did not perform the search for Higgs bosons with mass
exceeding 300~GeV.

In summary, we present a combination of CDF and D0 searches for the $gg\rightarrow H\rightarrow W^+W^-$
and $gg\rightarrow H\rightarrow ZZ$
processes and set an upper limit on $\sigma(gg\rightarrow H)\times \mathcal{B}(H\rightarrow W^+W^-)$ as a function
of $m_H$, assuming the standard model ratio of $\mathcal{B}(H\rightarrow W^+W^-)/\mathcal{B}(H\rightarrow ZZ)$.
 We compare these limits with the prediction of the minimal SM with a sequential
fourth generation of heavy fermions added on, and exclude at the 95\% C.L. the Higgs boson mass range
$124<m_H<286$~GeV, with an expected
exclusion of 120 -- 267 GeV.

\begin{center}
{\bf Acknowledgements}
\end{center}

We thank the Fermilab staff and the technical staffs of the
participating institutions for their vital contributions. 
This work was supported by  
DOE and NSF (USA),
CONICET and UBACyT (Argentina), 
CNPq, FAPERJ, FAPESP and FUNDUNESP (Brazil),
CRC Program, CFI, NSERC and WestGrid Project (Canada),
CAS and CNSF (China),
Colciencias (Colombia),
MSMT and GACR (Czech Republic),
Academy of Finland (Finland),
CEA and CNRS/IN2P3 (France),
BMBF and DFG (Germany),
Ministry of Education, Culture, Sports, Science and Technology (Japan), 
World Class University Program, National Research Foundation (Korea),
KRF and KOSEF (Korea),
DAE and DST (India),
SFI (Ireland),
INFN (Italy),
CONACyT (Mexico),
NSC(Republic of China),
FASI, Rosatom and RFBR (Russia),
Slovak R\&D Agency (Slovakia), 
Ministerio de Ciencia e Innovaci\'{o}n, and Programa Consolider-Ingenio 2010 (Spain),
The Swedish Research Council (Sweden),
Swiss National Science Foundation (Switzerland), 
FOM (The Netherlands),
STFC and the Royal Society (UK),
and the A.P. Sloan Foundation (USA).

\begin{table*}[htb]
\caption{\label{tab:cross_sections} The fourth-generation enhanced
$\sigma(gg\rightarrow H)$ listed in fb for the low-mass scenario
described in the text with $m_{d4} = 300$~GeV and $m_{d4} = 400$~GeV
along with the $\mathcal{B}(H\rightarrow W^+W^-)$ for Higgs masses
between 110 - 300~GeV.  The uncertainty on the predicted cross section
from variations in the PDF and factorization and renormalization scale
are also listed in percentage where these have been determined from
the MSTW 2008 90\% C.L. uncertainty and by modification of the scale
by factors of $1/2$ and 2, respectively. The inclusive uncertainties are listed here,
but the differential uncertainties in the jet categories are used in the calculation of the limits.}
\begin{ruledtabular}
\begin{tabular}{lcccccccc}
$m_H$&$\sigma(gg\rightarrow H)$ &$\sigma(gg\rightarrow H)$ &uncert.&uncert.&uncert.&uncert.& $BR(H\rightarrow W^+W^-)$&$BR(H\rightarrow W^+W^-)$\\
 (GeV)&$m_{d4}=300$~GeV&$m_{d4}=400$~GeV&PDF up(\%) &PDF down(\%) &$\mu$ up(\%) &$\mu$ down(\%) &$m_{d4}=300$~GeV&$m_{d4}=400$~GeV\\
\hline\hline
110& 12384  &12308 & 12 &-11 &12&-8& 0.028&0.028 \\
115 &10798  &10725  &12 &-11 &12&-8& 0.050&0.051\\
120 &9449.9 &9384.3 &12 &-11 &12&-8&0.083& 0.083\\
125 &8298.8 &8240.0 &12 &-12 &12&-8&0.13&0.13     \\
130 &7314.0 &7258.7 &12 &-12 &12&-8&0.19     &0.19  \\
135 &6465.1 &6414.2 &12 &-12 &12&-8&0.26     &0.26    \\
140 &5731.4 &5684.1 &13 &-12 &12&-8&0.35     &0.35     \\
145 &5094.6 &5050.4 &13 &-12 &12&-8&0.44     &0.44     \\
150 &4540.5 &4498.5 &13 &-12 &12&-8&0.55   &  0.55     \\
155 &4055.6 &4017.6 &13 &-12 &12&-8&0.68&0.68 \\
160 &3630.2 &3595.1 &13 &-13 &12&-8&0.85 &0.85 \\
165 &3253.7 &3220.7 &14 &-13 &12&-8&0.91 &0.91 \\
170 &2924.1 &2893.2 &14 &-13 &12&-8&0.89 &0.88 \\
175 &2633.9 &2604.4 &14 &-13 &12&-8&0.86 &0.86 \\
180 &2376.7 &2348.9 &14 &-13 &12&-8&0.83 &0.83 \\
185 &2147.2 &2121.5 &15 &-13 &12&-8&0.74 &0.74 \\
190 &1943.9 &1919.7 &15 &-14 &12&-8&0.69 &0.69 \\
195 &1763.2 &1740.2 &15 &-14 &12&-8&0.66 &0.66 \\
200 &1601.8 &1580.0 &15 &-14 &12&-8&0.65 &0.65 \\
210 &1328.1 &1308.4 &16 &-14 &12&-8&0.62 & 0.62 \\
220 &1107.7 &1089.6 &16& -15 &12&-8&0.60 &0.60 \\
230 &928.61 &912.21 &17& -15 &12&-8&0.59 &0.59 \\
240 &782.52 &767.44 &17 &-15 &12&-8&0.58 &0.58 \\
250 &662.60 &648.81 &18& -16 &12&-8&0.58 &0.58 \\
260 &563.53 &550.90 &19 &-16 &12&-8&0.57 &0.58 \\
270 &481.49 &469.93 &19 &-16 &12&-8&0.57 &0.57 \\
280 &413.24 &402.68 &20 &-17 &12&-8&0.58 &0.58 \\
290 &356.39 &346.53 &21& -17 &12&-8&0.58 &0.58 \\
300 &308.70 &299.71 &21 &-17 &12&-8&0.58 &0.58 \\
\end{tabular}
\end{ruledtabular}
\end{table*}

\begin{table*}[htb]
\caption{\label{tab:limits} The observed and median expected 95\% C.L. upper limits on
$\sigma(gg\rightarrow H)\times \mathcal{B}(H\rightarrow W^+W^-)$ for $m_H$ between
110~GeV and 300~GeV, obtained with the Bayesian and CL$_{\rm s}$ methods.
Also listed are the $\sigma(gg\rightarrow H)\times \mathcal{B}(H\rightarrow W^+W^-)$~predictions of the low-mass and the high-mass fourth-generation scenarios discussed
in the text for a fourth-generation down-type quark mass of 400~GeV.  All limits and predictions are presented in pb.}
\begin{ruledtabular}
\begin{tabular}{lcccccccc}
 $m_H$       &  \multicolumn{4}{c}{Bayes}& \multicolumn{2}{c}{CL$_{\rm s}$}&  4$^{\rm{th}}$ Gen &  4$^{\rm{th}}$ Gen \\
 $\left[{\rm GeV}\right]$ &Obs. &Ratio Low Mass & Exp.& Ratio Low Mass   &   Obs.             & Exp.          & Low Mass    & High Mass\\
 &&(Obs./4Gen)&&(Exp./4Gen)&&&\\\hline
110 & 	1.14& 3.08  &	1.14  & 3.30  &         &                    &        0.34 &  0.35	 \\
115 & 	1.03& 1.73  &	0.79  & 1.49  &    1.04 &	0.80         &        0.54 &  0.54	 \\
120 & 	1.01& 1.18  &	0.75  & 0.98  &    1.00 &	0.75         &        0.78 &  0.78	 \\
125 & 	1.19& 0.96 &	0.69  & 0.65  &    1.15 &	0.70         &        1.06 &  1.06	 \\
130 & 	1.21& 0.80 &	0.67  & 0.48  &    1.17 &	0.67         &        1.36 &  1.36	 \\
135 & 	1.31& 0.75 &	0.63  & 0.37  &    1.28 &	0.63         &        1.67 &  1.67	 \\
140 & 	1.50& 0.68 &	0.59  & 0.30  &    1.41 &	0.59         &        1.96 &  1.96	 \\
145 & 	1.02& 0.41 &	0.55  & 0.24  &    0.94 &	0.54         &        2.25 &  2.24	 \\
150 & 	0.75& 0.28 &	0.48  & 0.19  &    0.73 &	0.49         &        2.49 &  2.49	 \\
155 & 	0.8 & 0.27 &	0.41  & 0.15  &    0.76 &	0.42         &        2.74 &  2.74	 \\
160 & 	0.44& 0.14 &	0.30  & 0.095 &    0.42 &	0.30         &        3.06 &  3.06	 \\
165 & 	0.4 & 0.12 &	0.27  & 0.090 &    0.36 &	0.27         &        2.92 &  3.03	 \\
170 & 	0.34& 0.13 &	0.30  & 0.11  &    0.34 &	0.30         &        2.57 &  2.76	 \\
175 & 	0.34& 0.14 &	0.32  & 0.14  &    0.33 &	0.33         &        2.25 &  2.47    \\
180 & 	0.41& 0.20 &	0.35  & 0.18  &    0.40 &	0.35         &        1.96 &  2.17	 \\
185 & 	0.38& 0.24 &	0.36  & 0.22  &    0.37 &	0.37         &        1.57 &  1.76	 \\
190 & 	0.45& 0.34 &	0.37  & 0.28  &    0.46 &	0.38         &        1.32 &  1.48	 \\
195 & 	0.44& 0.37 &	0.40  & 0.34  &    0.43 &	0.41         &        1.15 &  1.30	 \\
200 & 	0.47& 0.47 &	0.41  & 0.39  &    0.47 &	0.43         &        1.02 &  1.15	 \\
210 & 	0.39& 0.49 &	0.41  & 0.50  &    0.39 &	0.42         &        0.81 &  0.94    \\
220 & 	0.42& 0.63 &	0.39  & 0.60  &    0.43 &	0.42         &        0.65 &  0.77	 \\
230 & 	0.30& 0.55 &	0.36  & 0.67  &    0.29 &	0.37         &        0.54 &  0.64	 \\
240 & 	0.37& 0.93 &	0.35  & 0.81  &    0.40 &	0.37         &        0.45 &  0.54	 \\
250 & 	0.24& 0.67 &	0.32  & 0.86  &    0.25 &	0.33         &        0.37 &  0.45	 \\
260 & 	0.23& 0.77 &	0.30  & 0.91  &    0.24 &	0.31         &        0.32 &  0.38    \\
270 & 	0.20& 0.71 &	0.28  & 1.04  &    0.21 &	0.29         &        0.27 &  0.33    \\
280 & 	0.21& 0.92 &	0.25  & 1.08  &    0.22 &	0.26         &        0.23 &  0.28    \\
290 & 	0.21& 1.05 &	0.24  & 1.20  &    0.21 &	0.24         &        0.20 &  0.24    \\
300 & 	0.17& 1.00 &	0.22  & 1.27  &    0.17 &	0.22         &        0.17 &  0.21    \\
\end{tabular}
\end{ruledtabular}
\end{table*}

\begin{thebibliography}{000}
\bibitem{dzHWWtau} \DZero Collaboration,
`` Search for the Standard Model Higgs boson in $ \mu + \tau_{had} \le 1$ jet final
state  with 7.3 fb$^{-1}$ of data'',
D0 Conference Note 6179.
\end{thebibliography}
\end{document}